# Observation of Supercurrent in PbIn-Graphene-PbIn Josephson Junction

Dongchan Jeong,[1] Jae-Hyun Choi,[1] Gil-Ho Lee,[1] Sanghyun Jo,[1, 2] Yong-Joo Doh,[3, *] and Hu-Jong Lee[1, 2, *]

[1]*Department of Physics, Pohang University of science of Technology, Pohang 790-784, Republic of Korea*
[2]*National Center for Nanomaterials Technology, Pohang 790-784, Republic of Korea*
[3]*Department of Display and Semiconductor Physics,*
*Korea University Sejong Campus, Jochiwon, Chungnam 339-700, Republic of Korea*
(Dated: January 28, 2011)

Superconductor-graphene-superconductor (SGS) junction provides a unique platform to study relativistic electrodynamics of Dirac fermions in graphene combined with proximity-induced superconductivity. We report observation of the Josephson effect in proximity-coupled superconducting junctions of graphene in contact with $Pb_{1-x}In_x$ ($x$=0.07) electrodes for temperatures as high as $T = 4.8$ K, with a large value of $I_c R_N$ ($\sim$255 $\mu$V). This demonstrates that $Pb_{1-x}In_x$ SGS junction would facilitate the development of the superconducting quantum information devices and superconductor-enhanced phase-coherent transport of graphene.

## I. INTRODUCTION

Properties of graphene, a monolayer honeycomb lattice of carbon atoms, have been investigated intensively ever since it was first discovered by the mechanical exfoliation.[1,2] To date, experimental and theoretical studies of graphene have mainly been focused on features arising from its unique band structure, described by the relativistic Dirac equation.[3-5] The chiral nature of charge carriers in graphene is revealed in transport measurements of the half-integer quantum Hall effect,[6-8] weak localization,[9,10] and Klein tunneling[11,12] etc.

A superconductor-graphene (SG) hybrid system, such as an SGS junction or an SG interface, provides an ideal platform to investigate the relativistic nature of Dirac fermions combined with superconductivity.[13,14] Instead of the retroreflection of carriers in an ordinary superconductor$-$normal-metal interface, an SG interface is theoretically predicted to show the specular reflection of chiral quasiparticle carriers.[14] The specular reflection of carriers near the Dirac point can be more easily observed if superconducting material with a large superconducting energy gap is employed for the electrodes because the effect of unexpected charge doping near the Dirac point is minimized. Moreover, phase-coherent transport in graphene of quasiparticles with the energy below the superconducting energy gap can be enhanced significantly due to the condensation of electrons in superconductor into a single macroscopic quantum state. This property can be applied for mesoscopic phase interferometer[15] of graphene. A large superconducting energy gap with a high transition temperature $T_c$ also provides a potential advantage for qubit application of graphene, where a Josephson element is operable at high temperatures. Yet, previous experimental studies have mainly been focused on SGS junctions with Al electrodes,[16-19] which have a relatively small superconducting gap energy of $2\Delta_{Al} \sim 250$ $\mu$eV (this value is smaller than the bulk value of $\sim$340 $\mu$eV) and low $T_c$ ($\sim$1 K). For Al-based SGS-junction devices measurements with high sensitivity at low temperatures below 1 K are required to overcome the external noise. Although other experimental results of SGS junctions employing W and Pt/Ta as superconducting electrodes have been reported,[20,21] either the supercurrent was not clearly seen[20] or it was achieved only after heavy annealing of the sample.[21] Thus, realizing SGS Josephson coupling employing a new superconducting material with a higher superducting energy gap, as in this study, can be regarded as significant progress in studies of superconducting-proximity effect in graphene.

We report on the fabrication and measurements of the SGS Josephson junctions employing $Pb_{0.93}In_{0.07}$ as the superconducting-electrode material with a higher $T_c$ of 7.0 K. The $Pb_{1-x}In_x$-based SGS junction exhibits the supercurrent up to $T\sim$4.8 K above the liquid-helium temperature, which is the highest operation temperature of an SGS junction reported to date. The junction response to the external magnetic or microwave field manifests the genuine Josephson characters expected by the theory.[22] Extensive studies on the electrical transport properties of our SGS junctions, which depend on the bias ($V$) and the back-gate voltage ($V_{BG}$), reveal the superconducting energy gap of $2\Delta_{PbIn} \sim 2.2$ meV. It is an order of magnitude higher than that of Al. The studies also reveal the subgap structures of differential conductance ($dI/dV$) induced by the multiple Andreev reflection.[23] Thus, SGS junctions consisting of $Pb_{1-x}In_x$ superconducting electrodes lead to the superior device performance over the previous Al-based SGS junctions.

## II. SAMPLE PREPARATION

We used $Pb_{1-x}In_x$ alloy instead of pure Pb as the superconducting-electrode material to minimize the granularity of electrodes while their $T_c$ remains almost intact. Both lead (Pb) and indium (In) are type-I superconductors with different superconducting transition temperatures ($T_{c,Pb} = 7.19$ K, $T_{c,In} = 3.40$ K). $Pb_{1-x}In_x$ alloy, however, makes a type-II superconductor. $T_c$ and critical magnetic field of which vary depending on the composition ratio.[24] The $Pb_{1-x}In_x$ film was thermally deposited at the rate of 0.7 nm/s in the base pressure of

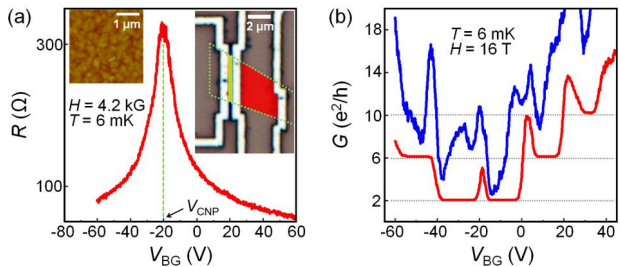

FIG. 1. (color online) (a) Gate-voltage dependence of the junction resistance obtained from the $Pb_{0.93}In_{0.07}$-based SGS junction in the normal state at $H$=4.2 kOe and at $T$=6 mK. The charge neutrality point ($V_{CNP}$=-20 V) is indicated by an arrow. Left inset: topographic AFM image of the $Pb_{0.93}In_{0.07}$ film showing granular morphology. Right inset: optical microscope image of the device. The dotted line denotes the boundary of the monolayer graphene. The narrow colored region indicates the 300-nm spacing of graphene layer between two $Pb_{0.93}In_{0.07}$ electrodes. The wide colored region of graphene layer used for the measurement of the quantum Hall plateaus, respectively. (b) Gate-voltage-dependent conductance of the narrow junction and the wide graphene sheet with $H$=16 T at $T$=6 mK. The dotted line corresponds to $G$=$\nu e^2/h$ ($\nu$ = 2, 6, 10 . . .).

$6 \times 10^{-7}$ Torr.

Atomic force microscope (AFM) measurement reveals that the $Pb_{1-x}In_x$ ($x$ = 0.04, 0.07, and 0.10) films exhibit a granular structure with the grain diameter of 300 $\sim$ 400 nm and the root-mean-square roughness of 9.5 nm [see the inset of Fig. 1(a)]. It should be noted that the $Pb_{1-x}In_x$ film on the graphene layer gets significantly thinner than the outer parts, which is probably caused by that the $Pb_{1-x}In_x$ is more mobile on graphene than directly on the oxidized Si substrate.[25] $T_c$ of $Pb_{1-x}In_x$ film on graphene is sensitive to its thickness, which in turn is responsible for the superconducting proximity effect in graphene. The thickness and the atomic composition of $Pb_{1-x}In_x$ electrodes were adjusted so as to obtain the highest value of $T_c$ of 7.0 K for $Pb_{1-x}In_x$ film. The contact resistance obtained from three-probe measurement configuration was in the range of $1-15 \ \Omega \cdot \mu m^2$ in most of our devices and it was insensitive to the temperature change.

Monolayer graphene was mechanically exfoliated from the natural graphite by using Scotch brand tape and transferred on a highly electron-doped silicon substrate covered with a 300-nm-thick oxidized-silicon wafer. The carrier density in the graphene was tuned by using the silicon substrate as a back gate. The 900-nm-wide electrode of $Pb_{0.93}In_{0.07}$/Au (200/10 nm in thickness) double layer was formed on top of a graphene flake with a spacing ($L$) of 300 nm between the electrodes. In this study, two sets of dilution fridges (Leiden Cryogenics Model MNK 126-500 and Oxford Instruments Model AST) were employed. The latter was used to measure the microwave response of the SGS junction. For the low-noise measure-ments, two-stage RC filters (cut-off frequency $\sim$ 30 kHz) and $\pi$ filters were connected in series with the measurement leads and a low-frequency ($\sim$ 13.3 Hz) conventional lock-in technique was adopted for the measurement of dynamic conductance $dI/dV$.

## III. RESULTS AND DISCUSSION

Fig. 1(a) shows the back-gate voltage ($V_{BG}$) dependence of the junction resistance at $T$ = 6 mK, obtained by quenching the superconductivity of $Pb_{0.93}In_{0.07}$ electrodes in an external magnetic field of $H$ = 0.42 T. It reveals the charge neutrality point ($V_{CNP}$) at $V_{BG}$ = -20 V. The carrier mobility and the mean-free path are estimated to be $\mu \sim 1400$ cm$^2$/V·sec and $l \sim 24$ nm, respectively, for $\Delta V_{BG}$ = -30 V, where $\Delta V_{BG} \equiv V_{BG}$ - $V_{CNP}$. The spacing of the graphene junction between two Al electrodes is an order of magnitude longer than the carrier mean-free path ($L > l$) in the graphene, which implies that all the data should be analyzed in terms of the proximity-junction model in a diffusive limit.

The single-layeredness of graphene is confirmed by the quantized Hall conductance measured in a high magnetic field of $H$ = 16 T. In a quantum-Hall regime, the conductance plateaus of $G$ = $\nu e^2/h$ with the quantized filling factor of $\nu$ = 2, 6. 10 are expected for the monolayer graphene,[1,2] where $h$ is the Planck's constant. In a two-probe configuration adopted in this study with the invasive superconducting contacts over the graphene layer, however, the Hall conductivity ($\sigma_{xy}$) is mixed with the longitudinal conductivity ($\sigma_{xx}$), which is dependent on the aspect ratio ($\zeta$ = $L/W$) of the graphene layer in the junction area.[26,27] Thus, conductance dips take place for the SGS junction with $\zeta$ = 0.05, while the neighboring junction with $\zeta \sim 0.9$ on the same graphene layer exhibits the quantized conductance plateaus, as shown in Fig. 1(b).

The resistive transition of the $Pb_{0.93}In_{0.07}$ electrodes shows an onset of the superconductivity at $T_{c,onset}$ = 7.15 K, while the electrodes become fully superconducting below $T_c$ = 7.0 K [see the inset of Fig. 2(a)]. The Josephson coupling is established through the graphene layer in the SGS junction below $T_c$ of $Pb_{0.93}In_{0.07}$ electrodes. In Fig 2(a), the current-voltage ($I$-$V$) curve obtained at the base temperature ($T$ = 6 mK) shows a sharp switching from a dissipationless to a resistive state at a critical current ($I_c \sim 3.0 \ \mu A$). A reverse switching occurs at a retrapping current ($I_R \sim 0.8 \ \mu A$), which yields the clear hysteretic behavior. The ambient magnetic field was canceled by applying a compensation field of $H$ = 1.84 Oe, at which $I_c$ was maximized.

Progressive evolution of the $I$-$V$ characteristics with temperature is displayed in Fig. 2(b). where the bias current is swept from negative to positive polarity. The asymmetry in $I$-$V$ curves represents the occurrence of hysteresis. With increasing $T$, $I_c$ becomes smaller and the hysteresis is reduced. It is noted that $I_c$ remains finite for $T$ as high as 3.83 K. After the thermal recycling (*i.e.*, switching the cryostat from Leiden Cryogenics Model MNK 126-500 to Oxford Instruments Model AST), $I_c$ became observable up to 4.8 K above the liquid-helium temperature [see the Supplementary Information].[25] The corresponding critical current density, $J_c(0) \sim 9.4 \times 10^{-3}$ A/cm, is a record high obtained from SGS junctions to date.

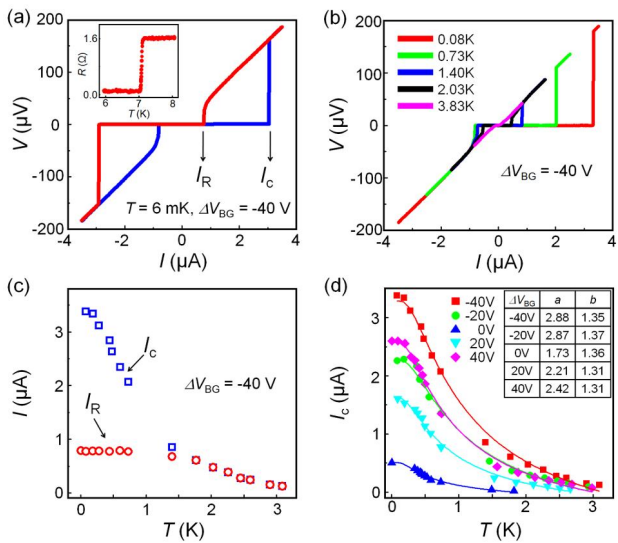

FIG. 2. (color online) (a) Current-voltage ($I$-$V$) characteristic curve measured at $\Delta V_{BG}$=-40 V with increasing and decreasing bias current. The critical ($I_c$) and the retrapping ($I_R$) currents are indicated. Inset: resistance vs temperature ($R$-$T$) curve of a Pb$_{1-x}$In$_x$ electrode, revealing $T_c$=7.0 K. (b) $I$-$V$ curves with five different temperatures at $\Delta V_{BG}$=-40 V. (c) $T$ dependencies of $I_c$ and $I_R$ for $\Delta V_{BG}$=-40 V. (d) $I_c$-$T$ curves for different back-gate voltages. The solid lines are the theoretical fits explained in the text.

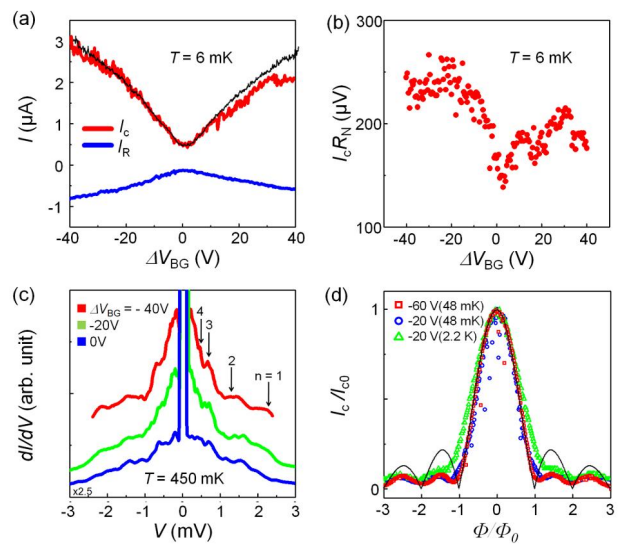

FIG. 3. (color online) (a) Gate-voltage dependence of $I_c$ (upper curve) and $I_R$ (lower curve) at $T$=6 mK with bias current swept from negative to positive polarity. (b) $I_c R_N$ product as a function of the back-gate voltage. (c) $dI/dV$ vs $V$ curves with varying $\Delta V_{BG}$=0, -20, -40 V from bottom to top. The conductance with $\Delta V_{BG}$=0 V is magnified by 2.5 times for the sake of clarity. The arrows indicate the subgap structures corresponding to the voltage of $V_n$=$2\Delta_{PbIn}/ne$ ($n$=1, 2, 3, 4). (d) The normalized critical current ($I_c$) as a function of magnetic flux ($\Phi$) measured at $\Delta V_{BG}$=-60 V and -20 V for $T$ = 48 mK and -20 V for $T$=2.2 K. The maximum critical currents ($I_{c0}$) are $I_{c0}$=5.8, 2.8, 0.8 $\mu$A for $\Delta V_{BG}$ ($T$)=-60 (0.048 K), -20 (0.048 K) and -20 (2.2 K) V, respectively. The black curve represents a calculation result of an ideal Fraunhofer pattern (see text).

Typical temperature dependencies of $I_c$ and $I_R$ are displayed in Fig. 2(c) for $V_{BG}$ = -40 V. $I_c$ decreases rapidly with increasing temperature but $I_R$ remains almost constant up to $T$ = 1.77 K, at which the hysteresis disappears (or $I_c$ becomes equal to $I_R$). Similar $T$ dependence of $I_c$ is obtained for different gate voltages, as illustrated in Fig. 2(d). The overall $I_c$-$T$ curves will fit the diffusive Josephson-junction behavior in a long junction regime;[28]

$$eI_c R_N = aE_{Th}\left[1 - b\exp\left(\frac{-aE_{Th}}{3.2k_B T}\right)\right] \quad (1)$$

where $E_{Th}$ is the Thouless energy and $a$ and $b$ are fitting parameters. From the $R$ - $V_{BG}$ curve in Fig 1(a), $E_{Th}$(=$\hbar D/L^2$; $\hbar$≡$h/2\pi$) is estimated to be about 90 $\mu$eV for $V_{BG}$ sufficiently away from the CNP ($\Delta V_{BG}$=-40 V) and is weakly dependent on $\Delta V_{BG}$. Here, $D$ (= $v_F l/2$; $v_F$ is the Fermi velocity) is the diffusion constant of graphene. In comparison with the theoretical expectation of $a$ = 10.8 and $b$ = 1.30, in a long junction limit of $E_{Th}/\Delta_{PbIn} \rightarrow 0$, the best-fit values of parameters $a$ and $b$ of our device turn out to be 1.7 - 2.9 and ~1.3, respectively, with $E_{Th}/\Delta_{PbIn}$=0.083. Our PbIn-based SGS junction, however, corresponds to an intermediate regime between the long and short junction limits (see the discussion below), where a smaller value of $a$ is expected from the theoretical calculation of Ref. [28]. This may lead to the reduction of the parameter $a$ in our junction. We note that the onset temperature ($T^*$) of a finite $I_c$ is strongly dependent on the gate voltage, resulting in the highly increased $T^*$ and much larger $I_c(T = 0)$ as well for the gate voltages further away from $V_{CNP}$. This is attributed to a competition between the Josephson coupling energy ($E_J$=$\hbar I_c/2e$) and thermal fluctuations. Due to a small value of $I_c$ near $V_{CNP}$, the supercurrent is vulnerable to the thermal fluctuation, which results in the reduction of $T^*$.

Gate-voltage dependencies of $I_c$ and $I_R$ at $T$ = 6 mK are displayed in Fig. 3(a), the magnitude of which increases monotonously with $\Delta V_{BG}$. As observed previously in SGS junctions[16] the $V_{BG}$ dependence of $I_c$ correlates with the $V_{BG}$ dependence of the normal-state conductance $G(V_{BG})$ represented by the black curve. The significant difference between the two ($I_c > I_R$) indicates that the hysteretic $I$-$V$ curves prevail over the whole $V_{BG}$ even including $V_{CNP}$. Similar hysteresis is observed in various proximity-coupled Josephson junctions consisting of normal metals,[29] semiconductor nanowires,[30,31] and carbon nanotubes.[32] In a resistively and capacitively shunted junction (RCSJ) model,[22] which is a qualitative model for the Josephson junction, the hysteresis can be explained by the presence of a finite junction capacitance. Although the geometric capacitance ($C$) is supposed to be negligible in the SGS junction, the effective capacitance ($C_{eff}=\hbar/R_N E_{Th}$) can be suggested by replacing the quasiparticle characteristic time ($\tau = R_N C$) by the quasiparticle diffusion time in graphene between the electrodes ($\tau_D=\hbar/E_{Th}$).[33] where $R_N$ is the junction resistance in the normal state. Because of the gate-voltage independence of $E_{Th}$, $C_{eff}$=88 fF at $\Delta V_{BG}$=-40 V is inversely proportional to $R_N$. The quality factor for the junction, which is defined as $Q=(2eI_cR_N/E_{Th})^{1/2}$ in terms of $C_{eff}$, varies in the range of $Q$=1.8 - 2.3, depending on $\Delta V_{BG}$. The approximate expression of $Q \sim I_c/I_R$ yields a comparable result of 3.4 - 4.5, as can be estimated from Fig. 3(a). This supports our interpretation of the appearance of the hysteresis in terms of the effective capacitance $C_{eff}$.

The $I_cR_N$ product, which is a figure of merit for a Josephson junction, is given in Fig. 3(b) as a function of $\Delta V_{BG}$. The maximum $I_cR_N$ value reaches 255 $\mu$V for $\Delta V_{BG}$ = -40 V and it reduces to 145 $\mu$V at $V_{CNP}$. The ratio of the maximum $I_cR_N$ product to the superconducting gap energy turns out to be $eI_cR_N/\Delta_{PbIn}$ = 0.23, where the value of $\Delta_{PbIn}$ = 1.1 meV determined by the multiple Andreev reflection (to be discussed below) is adopted. Combining with the ratio of $E_{Th}/\Delta_{PbIn}$ = 0.083 obtained from the $I_c - T$ curves in Fig. 2, the value of $eI_cR_N/\Delta_{PbIn}$ indicates that our PbIn-based SGS junction is in the intermediate regime between the short ($E_{Th}/\Delta_{PbIn}$ > 1) and the long ($E_{Th}/\Delta_{PbIn}$ < 0.01) junction limits. According to Ref. [28], the ratio of $E_{Th}/\Delta_{PbIn}$ = 0.083 of our device should correspond to $eI_cR_N/\Delta_{PbIn}$ ∼0.5 in the zero-temperature limit. The reduction of $I_cR_N$ of our device below the expected value may be attributed to the suppression of $I_c$ due to the thermal fluctuations and/or incomplete filtering of the external noise. It should be noted that the observed variation of $I_cR_N$ with $\Delta V_{BG}$ can be attributed to $V_{BG}$ dependence of $E_{Th}$, except for the region of $V_{BG}$ close to the CNP ($|\Delta V_{BG}|$ < 10 V) where the carrier density strongly fluctuates due to the presence of electron-hole puddles.

When the highly transparent contact forms at the normal-metal−superconductor (NS) interface, the Andreev reflection process[23,34] takes place, where an incident electron from the normal metal with energy below $2\Delta_{PbIn}$ is reflected as a hole and a Cooper pair propagates into the superconductor, or vice versa. In an SNS junction a quasiparticle in N, accelerated by potential difference $eV_n$, can be Andreev-reflected multiple times until its energy exceeds $2\Delta_{PbIn}$. This multiple Andreev reflection (MAR) is responsible for the subgap conductance peaks occurring at voltages of $V_n=2\Delta_{PbIn}/ne$, where $n$ is an integer, as shown in Fig. 3(c). The corresponding superconducting energy gap of $Pb_{0.93}In_{0.07}$ electrodes is estimated to be $2\Delta_{PbIn}$ = 2.2 meV, which is comparable to the bulk value of 2.7 meV. The subgap structures are evident at the constant $V_n$ values for different gate voltages and become clearer for $V_{BG}$ away from $V_{CNP}$. This is attributed to a longer phase coherence length of quasiparticles in a higher-doped region.[35]

A direct evidence of the genuine Josephson coupling through the graphene layer, rather than any artifact such as an inadvertent formation of micro-bridge of diffused $Pb_{1-x}In_x$ electrodes, is provided by a periodic modulation of $I_c$ in a perpendicular magnetic field, $i.e.$, so-called Fraunhofer pattern.[22] For a conventional Josephson junction, $I_c(H)$ exhibits local minima for integer magnetic flux quanta through the junction area except for $H = 0$. Figure 3(d) shows the magnetic field dependence of $I_c$ for different $V_{BG}$ and $T$. At very low temperature of $T$=48 mK, the periodic oscillations of $I_c(H)$ is clearly manifested with the $H$ period of $H_0=\Phi_0/A=\Phi_0/[(L + 2\lambda)W]$∼2.8 Oe, where $\Phi_0=h/2e$ is the magnetic flux quantum, $W$ is the junction width, and $\lambda$ is the London penetration depth of the superconducting $Pb_{0.93}In_{0.07}$ electrodes.[19] The theoretical prediction follows the relation[22] of $I_c(H)=I_c(0)\sin[2\pi(\Phi/\Phi_0)]/(\Phi/\Phi_0)$, which is in good agreement with the experimental results at low magnetic flux of $\Phi=HA$. The $I_c(H)$ modulation is observable even at the elevated temperature of $T$=2.2 K. The tunable Josephson coupling with the external magnetic field is very important for superconducting device applications such as SQUID's[36] or mesoscopic phase interferometers[15] of graphene.

Another crucial experimental evidence of the genuine Josephson effect is provided by the microwave response of the junction. A Josephson junction, irradiated by a microwave of frequency $f$, exhibits quantized voltage plateaus (or Shapiro steps) in the $I$-$V$ curve at discrete bias voltages of $V_n = nhf/2e$,[22] where $h$ is the Planck's constant. Figure 4(a) shows $I$-$V$ curves obtained from the $Pb_{0.93}In_{0.07}$-based SGS junction under the microwave irradiation of $f$=6 GHz with varying $\Delta V_{BG}$. The voltage interval between neighboring Shapiro steps is obtained to be $V_n$=12.4 $\mu$V, which is irrespective of $\Delta V_{BG}$ and precisely corresponds to the theoretical expectation of $V_n=hf/2e$. The proportionality relation between $\Delta V_n$ and $f$ with the slope of $h/2e$=2.07 $\mu$V/GHz is also confirmed in the frequency range of $f$=6−21 GHz as shown in Figs. 4(c) and 4(d). The widths of the Shapiro steps in the current axis is quasi-periodically modulated with the microwave power ($P^{1/2}$), as shown in the inset of Fig. 4(a), which resembles the Bessel-function-like behavior.[30] It is surprising to observe the quantized voltage plateaus at the elevated temperature up to $T$ = 4.8 K in Fig. 4(b), although the step edges are rounded by thermal fluctuations. The $dI/dV$-$V$ curves in the inset indicate that the Shapiro steps are almost vanishing only at $T$ = 5.4 K.

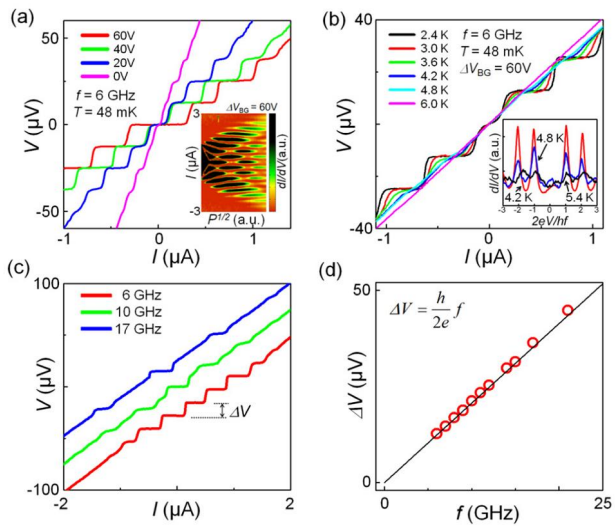

FIG. 4. (color online) (a) Gate-voltage dependent $I$-$V$ curves under the irradiation of a microwave of $f$=6 GHz at $T$=48 mK. Inset: color-coded plot of $dI/dV$ as a function of $I$ and square root of the microwave power ($P^{1/2}$) at $\Delta V_{BG}$=60 V. (b) $I$-$V$ curves measured at various temperatures under the irradiation of a microwave with $f$=6 GHz at $\Delta V_{BG}$=60 V. Inset: $dI/dV$-$V$ curve obtained at $T$=4.2, 4.8, and 5.4 K. The voltage was normalized by $hf/2e$ to clarify the existence of the Shapiro steps at $T$=4.2 and 4.8 K. (c) $I-V$ characteristic curves at 48 mK and $\Delta V_{BG}$ = -60 V with varying external microwave of frequency $f$ = 6 GHz, 10 GHz and 17 GHz from bottom to top. $\Delta V$ is the voltage interval between two different neighboring Shapiro steps. The curves are shifted for clarity. (d) The voltage interval $\Delta V$ as a function of microwave frequency $f$. The solid line shows a good agreement with the ac Josephson relation $\Delta V = hf/2e$.

## IV. SUMMARY

In summary, we demonstrate the realization of $Pb_{1-x}In_x$-based SGS Josephson junctions. Temperature dependence of the critical current and the value of $I_cR_N$ product of our system are well understood by the diffusive Josephson junction model. The dc and ac Josephson effects through the graphene layer are sustained over the liquid helium temperature. Furthermore, the experimental play ground for the graphene-based Josephson junction will be extended to much wider ranges of temperature, bias voltage, and magnetic field to explore the relativistic and superconducting phenomena with a reduction of unintended doping effect near the charge-neutrality point. Previous requirement of ultra-low temperatures ($T \ll 1$ K) to observe the supercurrent in the conventional Al-based SGS junctions can be resolved by employing the $Pb_{1-x}In_x$-based junctions, which can be readily generalized for other nanostructures, including carbon nanotubes and semiconductor nanowires. Ultimately, the superior Josephson-junction properties of $Pb_{1-x}In_x$-based junctions would leads to a very crucial step toward the application to graphene-based quantum information devices and gate-tunable Josephson devices.[37–39]

## ACKNOWLEDGMENTS

This work was supported by National Research Foundation of Korea (NRF) (for HJL) through Acceleration Research Grant R17-2008-007-01001-0 and Grant 2009-0083380, and (for YJD) through Grant 2010-0008450 (Basic Science Research Program) funded by the Ministry of Education, Science and Technology.

# Supplementary information accompanying "Observation of Supercurrent in PbIn-Graphene-PbIn Josephson Junction"

Dongchan Jeong,[1] Jae-Hyun Choi,[1] Gil-Ho Lee,[1] Sanghyun Jo,[1, 2] Yong-Joo Doh,[3, *] and Hu-Jong Lee[1, 2, *]

[1]*Department of Physics, Pohang University of science of Technology, Pohang 790-784, Republic of Korea*
[2]*National Center for Nanomaterials Technology, Pohang 790-784, Republic of Korea*
[3]*Department of Display and Semiconductor Physics,*
*Korea University Sejong Campus, Jochiwon, Chungnam 339-700, Republic of Korea*
(Dated: January 28, 2011)

Figure S1 shows the AFM images obtained from the graphene devices with (a) PbIn(100 nm), (b) PbIn/Au(200/5 nm), and (c) Ti/PbIn/Au(5/120/10 nm) electrodes. The profile of the hight of the 100-nm-thick PbIn electrode [see Fig. S1(d)] indicates that the thickness of the electrode on top of the graphene layer is reduced almost by half. The thickness reduction is relatively moderate for the 200-nm-thick PbIn electrode [see Fig. S1(e)] but the reduction rate is still ∼ 20%. Since a thinner superconducting film can exhibit a lower superconducting transition temperature and a reduced superconducting energy gap, thinning of PbIn electrodes may result in the suppressed superconducting proximity effect. But, when a Ti adhesion layer was inserted between graphene and PbIn no thickness reduction of the electrode was observed [see Fig. S1(f)]. This is understood by that the PbIn-alloy grains on the graphene layer are more mobile than directly on the SiO₂ layer of the substrate, while the difference in the mobility is alleviated on the common Ti contact layer. Diffusion of materials during thermal deposition is common for metallic elements with low melting temperatures. The thermal diffusion effect can be avoided by quench-condensing materials on substrates cooled by liquid helium.[1] Although the thickness of PbIn layers become more uniform with the Ti contact layer, the superconducting proximity coupling was seldom realized in the devices with Ti/PbIn electrodes. The curve of normalized conductance versus bias voltage at $T = 4.3$ K, as shown in Fig. S2, reveals that the sub-gap conductance enhancement due to the Andreev reflection is almost negligible in a device with Ti/PbIn(5/200 nm) electrodes, while the conductance is enhanced by 46% at zero bias in a device with PbIn(200 nm) electrodes. Thus we focus on the devices with PbIn electrodes without Ti contact layers. Au capping was made to all devices to avoid oxidation of the PbIn electrodes. The S/G transmission was irrelevant to the presence of the capping Au layer.

Figure S3 shows the basic device characteristics taken for the second cooling (in Oxford Instruments Model AST) after the thermal recycling. some in comparison with the data for the first cooling (in Leiden Cryogenics Model MNK 126-500). Figure S3(a) shows that, after thermal recycling, the charge neutrality point shifts from -20 V to -3.5 V, closer to the zero backgate voltage. The corresponding mobility is enhanced from 1,400 cm$^2$/V·sec to 1,700 cm$^2$/V·sec at $\Delta V_{BG}$ = -30V ($\Delta V_{BG} = V_{BG} - V_{CNP}$). We believe that the improvement of the transport characteristics after thermal recycling is owing to the possible release of adatoms or PMMA remnants during the fabrication. One notes that the charge neutrality point and mobility are known to be improved by the thermal annealing of graphene[2] in the atmosphere of Ar/H$_2$ mixture gas. Changes of the characteristics in our device are not abrupt as by intentional thermal annealing, but the thermal recycling up to room temperature and prolonged pumping of the device may have led to the gradual improvement. Improved normal-state properties of the graphene layer are also reflected in its superconducting state. Figure S3(b) shows the gradual change of the $I - V$ characteristics at $V_{BG}$ = 60V with increasing temperature. One notes that the Josephson supercurrent is maintained up to significantly higher temperatures in the second cooling. In the inset of Figure S3(b), the thermally smeared supercurrent feature or the conductance enhancement near zero bias is seen at 4.8 K and fully disappears only at 5.4 K. In contrast, the retrapping current ($I_R$) is less sensitive to the temperature change, only starting to decrease at temperature near 1.8 K similar to the observation for the first cooling. The improvement of the temperature dependence of the transport properties in graphene can be directly examined if the critical current ($I_C$) is normalized by the maximum critical current at the base temperature ($I_{C0}$). In Figure S3(c), the $I_C/I_{C0}$ at the same gate voltage away from the charge neutrality point ($\Delta V_{BG} = V_{BG} - V_{CNP1,2}$ = -40 V) exhibits the clear difference in the temperature dependence. $I_C$ decreases more gradually than for the first cooling. Even with this change, the diffusive character of the Josephson junction does not change because the mean free path ($l \sim 29$ nm) is still shorter than the junction distance (L). The $I_C$ vs $T$ curve in Figure S1(d) remains to well fit the long and diffusive junction model mentioned in the main text. The fitting parameters $a$ = 2.83 and $b$ = 1.25 at $\Delta V_{BG}$ = 60 V are close to the parameters obtained for the first cooling in Leiden Cryogenics Model MNK 126-500.

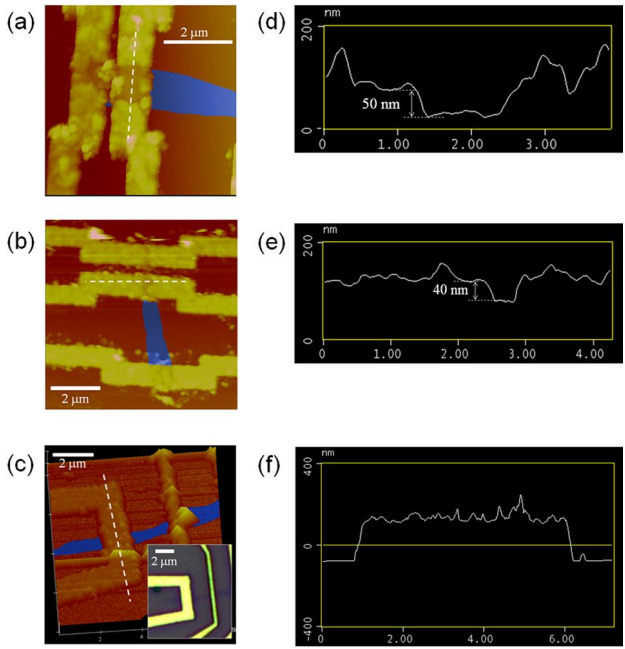

FIG. S1. (color online) False-color AFM images of graphene devices with (a) PbIn(100 nm), (b) PbIn/Au(200/5 nm) and (c) Ti/PbIn/Au(5/120/10 nm) electrodes. The graphene layer is emphasized using blue color. (d-f) The height profiles along the dotted lines corresponding to the images in (a-c), respectively.

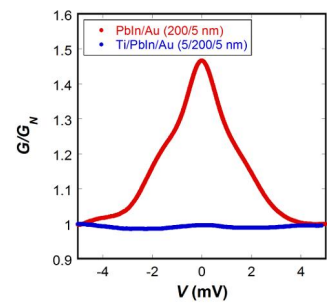

FIG. S2. (color online) Conductance versus bias voltage curves at $T = 4.3$ K, obtained from different tow graphene devices with PbIn/Au (red line) and Ti/PbIn/Au (blue) electrodes. The conductance (G) was normalized with the normal-state one ($G_N$) at high bias, where $G_N = 3.4$ mS (red line) and 2.7 mS ( blue).

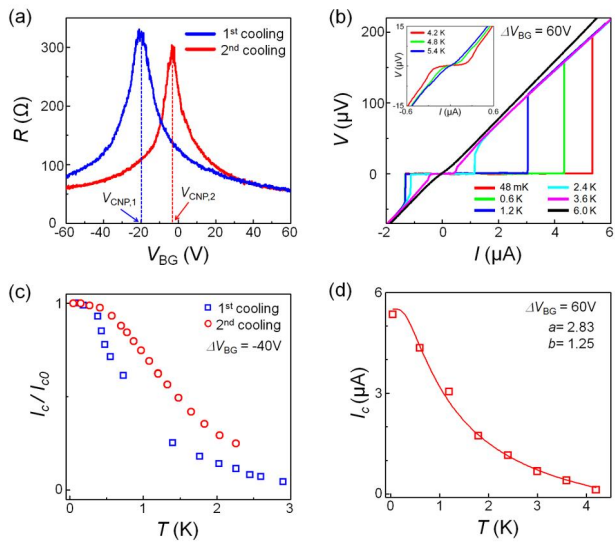

FIG. S3. (color online) (a) $I - V$ characteristic curves at 48 mK and $\Delta V_{BG}$ = -60 V with varying external microwave of frequency $f$ = 6 GHz, 10 GHz and 17 GHz from bottom to top. $\Delta V$ is voltage difference between two different neighboring Shapiro steps. The curves are shifted for clearness. (b) The voltage difference $\Delta V$ (red dots) as a function of microwave frequency $f$. The black solid line shows the good agreement with ac Josephson relation $\Delta V = hf/2e$.

---

* To whom correspondence should be addressed. E-mail: (Y.J.D.) yjdoh@korea.ac.kr; (H.J.L.) hjlee@postech.ac.kr